\documentclass[12pt,a4paper]{article}
\usepackage{setspace}
\usepackage[pdftex,colorlinks=true]{hyperref}
\usepackage{amssymb,enumitem,booktabs,xcolor,microtype,paralist,fullpage, makecell, longtable, bm, orcidlink, dsfont}
\usepackage[top=0.94in, bottom=0.94in, left=1in, right=1in]{geometry}
\usepackage{natbib}
\usepackage{url}
\definecolor{darkblue}{rgb}{0,0,.6}
\hypersetup{citecolor=darkblue,linkcolor=darkblue,urlcolor=darkblue}
\usepackage{amsmath}
\usepackage[font=small]{caption}
\usepackage{amsfonts}
\usepackage{epsfig}
\usepackage{graphics}
\usepackage{subcaption}
\usepackage{palatino,eulervm}
\usepackage{graphicx}
\usepackage{lscape}
\usepackage{rotating}
\usepackage[linewidth=1pt]{mdframed}
\allowdisplaybreaks[4]
\DeclareMathOperator*{\argmin}{arg\,min}

\newsavebox\CBox
\def\textBF#1{\sbox\CBox{#1}\resizebox{\wd\CBox}{\ht\CBox}{\textbf{#1}}}

\usepackage[font=small]{caption}

\providecommand{\U}[1]{\protect\rule{.1in}{.1in}}
\renewcommand{\baselinestretch}{1.2}
\graphicspath{{plots/}}

\usepackage{amsthm,thmtools}
\usepackage{mathrsfs}

\makeatletter
\def\th@newremark{\th@remark\thm@headfont{\bfseries}}
\makeatletter
\theoremstyle{newremark}

\declaretheoremstyle[spaceabove=6pt, spacebelow=6pt, headfont=\bfseries, notefont=\mdseries, notebraces={(}{)}, bodyfont=\normalfont, postheadspace=0.5em]{mystyle}

\newcommand{\Rlogo}{\protect\includegraphics[height=1.8ex,keepaspectratio]{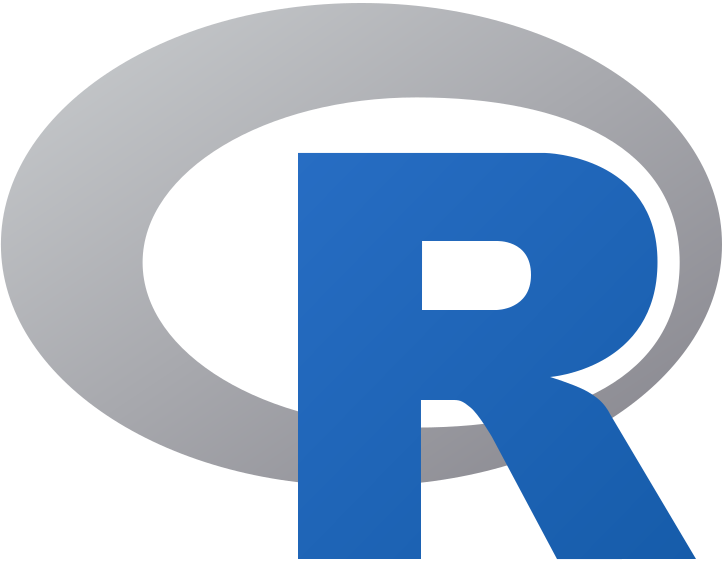}}

\begin{document}

\title{\Large\bf Forecasting Australian Electricity Generation by Fuel Mix}
\author{\normalsize  Han Lin Shang \orcidlink{0000-0003-1769-6430} \qquad Lin Han \orcidlink{0000-0002-4107-9531} \qquad Stefan Tr\"uck \orcidlink{0000-0002-5302-3850} \\
\normalsize Department of Actuarial Studies and Business Analytics \\
\normalsize Macquarie University \\
}
\date{}

\maketitle

\begin{abstract}
Electricity demand and generation have become increasingly unpredictable with the growing share of variable renewable energy sources in the power system. Forecasting electricity supply by fuel mix is crucial for market operation, ensuring grid stability, optimizing costs, integrating renewable energy sources, and supporting sustainable energy planning. We introduce two statistical methods, centering on forecast reconciliation and compositional data analysis, to forecast short-term electricity supply by different types of fuel mix. Using data for five electricity markets in Australia, we study the forecast accuracy of these techniques. The bottom-up hierarchical forecasting method consistently outperforms the other approaches. Moreover, fuel mix forecasting is most accurate in power systems with a higher share of stable fossil fuel generation. 

\vspace{.05in}
\end{abstract}

\newpage
\def\spacingset#1{\renewcommand{\baselinestretch}{#1}\small\normalsize} \spacingset{1}
\spacingset{1.58} 

\section{Introduction}\label{sec:intro}

Over the last decade, electricity markets around the world have been transitioning from predominantly fossil fuel-based generation to increasingly higher shares of renewable energy sources. This shift has introduced significant challenges, including greater variability and uncertainty in supply, increased price volatility, and reduced system inertia \citep{Hirth2013,Neuhoff2023,Rosales2024}. As variable renewable energy sources account for a growing share of the power mix, electricity demand and generation have become increasingly challenging to forecast. Traditional market structures, designed for controllable generation, often struggle to accommodate the characteristics of wind and solar power. As a result, there is a growing need for enhanced flexibility, improved forecasting, and market reforms to ensure reliability, efficient dispatch, and adequate investment signals in an evolving power system \citep{IEA2020}. 

This study contributes to the literature by introducing two statistical approaches, namely forecast reconciliation and compositional data analysis, for short-term forecasting of electricity supply across different fuel types. Energy consumption and supply forecasting is crucial for system planning, but also for the successful operation of wholesale electricity markets, where demand and supply must be balanced with extreme precision \citep{Weron2006}. High forecast accuracy contributes to significant improvements in energy management to achieve better grid security and reliability via mechanisms, such as load shifting, supply curtailment and energy storage. For energy market operators precise forecasts of electricity generation and demand are vital for the reliable management of the market, while for market participants these forecasts help to develop adequate operating and trading strategies \citep{Maciejowska2021, Babalhavaeji2023, Mesa2023, zheng2023hybrid, xia2025}.

Substantial work has been carried out in the existing literature focusing on developing various demand forecasting models, including both short-term (i.e., intraday, daily or weekly) forecasting that is important for wholesale trading, system reliability tasks and dispatch scheduling and longer-term electricity consumption or supply forecasting that supports strategic system planning and grid expansion. Commonly used approaches include statistical models, especially time series models \citep[see, e.g.,][]{shang2013functional,amini2016arima,spiliotis2020cross}, machine learning and deep learning techniques \citep[see, e.g.,][]{ruiz2018energy,cai2019day, bedi2019deep, Babalhavaeji2023}, and hybrid methods \citep[see, e.g.,][]{jurado2015hybrid,shahid2021novel,zheng2023hybrid}.

In short-term electricity consumption and supply forecasting \citep{de2013electricity,al2018two,sekhar2023robust,grandon2024electricity}, the most commonly used features include the historical consumption series, seasonal patterns such as weekdays and holidays, and weather features such as temperature, solar radiation and wind speed. In the longer term, electricity consumption and supply can be impacted more by changes in macroeconomic factors such as GDP and population, weather and climate factors, and supply and prices in fuel markets \citep{azadeh2008annual,kankal2011modeling,wu2018using}. A full review of energy forecasting techniques is outside the scope of this work, but a comprehensive review can be found in \cite{vom2020data}. A limited amount of forecasting literature has put a focus on the supply side. In particular, forecasting the specific fuel generation and fuel mix of the energy supply are lacking in attention. The focus in this area has been on the forecast of variable renewable energy (VRE) generators, i.e., wind and solar \citep[see, e.g.,][]{lee2017forecasting, shahid2021novel, Mesa2023, zheng2023hybrid, Cui2025}. For a recent review on forecasting methods for short-term wind power generation, see, e.g., \cite{Arslan2024}.   

Despite the significant gap in the current literature focusing on this topic, forecasting electricity supply from individual technologies and the generation fuel mix can provide substantial benefits across multiple dimensions:
\begin{asparaenum}
\item [1)] Fuel mix forecasting is crucial for the efficient and secure operation of electricity markets, as it informs real-time dispatch decisions, system balancing, and reserve planning. Recently, this task has become significantly more challenging due to the decreasing share of fossil-fuel fired dispatchable generation and the increasing share of intermittent renewable energy sources. These changes have introduced greater uncertainty and complexity into short-term operational planning that are also indicated by recent reliability issues in wholesale electricity markets \citep{Csereklyei2021,Rangarajan2025}.
\item [2)] Forecasting the supply mix is important for price discovery in wholesale electricity markets, as different generation technologies have very different short-run marginal costs \citep{Nazifi2021,Borenstein2022}. Anticipating which technologies will be dispatched helps market participants better predict price levels and bidding strategies.
\item[3)] It enhances market participants' risk management strategies by providing insights into fuel supply dynamics. This is of significant importance not only for VRE generators, which are subject to substantial volatility in output, but also for flexible generators, whose production is contingent upon VRE supply levels.  As pointed out by \cite{hodge2018combined}, the value of forecasting energy supply and storage increases significantly with higher penetration of solar and wind. 
\item[4)] It contributes to improved accuracy of aggregate consumption and supply forecasts, given that different fuel types typically exhibit distinct supply patterns. Consequently, accurate forecasts of individual fuel sources can refine overall supply projections. 
\item[5)] The forecast of future fuel mix also guides investors in their decision-making as their potential revenue is highly related to the amount of electricity they supply.
\item[6)] The energy production sector remains a principal contributor to carbon emissions \citep{EnergyTransitionsCommission2023}. Given the direct correlation between energy supply and emissions \citep{qader2021forecasting, aryai2023day}, robust forecasting of energy supply plays a critical role in emission estimation, control, and strategic planning for achieving net-zero targets.
\end{asparaenum}

In this study, we aim to model and forecast the generation fuel mix in the Australian national electricity market (NEM). We propose two approaches that adhere to a hierarchical structure, ensuring that the forecasts for each component of the generation fuel mix aggregate coherently to the total forecast. The Australian NEM is one of the world's longest interconnected power systems \citep{ignatieva2016, Regulator2023}. It employs a diverse range of fuel sources to generate electricity, containing both traditional fossil fuels such as coal and gas, as well as renewable energy sources including wind, solar, and hydropower. The NEM is undergoing a significant energy transition aimed at achieving net-zero emissions by 2050 \citep{AEMO2024, Ming2024}. This transition involves a gradual phasing out of traditional fossil fuels, which are being increasingly replaced by renewable energy sources. The incorporation of VREs introduces a degree of uncertainty and volatility into the electricity supply, leading to significant variations in availability and generation. Consequently, the generation mix within the NEM exhibits continuous fluctuations, reflecting both the long-term shift towards renewables and the short-term volatility associated with the growing presence of VREs.

In energy modeling, there are practical applications requiring forecasts of multiple time series organized in a hierarchical or grouped structure based on various attributes, such as fuel mix type. This has led to the necessity for reconciling forecasts across the hierarchy, ensuring that the forecasts sum adequately across all levels of the hierarchy. We evaluate and compare two approaches to forecast fuel mix, namely, hierarchical time series forecasting \citep[see][for a comprehensive review]{AHK+24} and compositional data analysis (CoDa) \citep[see][for a comprehensive review]{Greenacre21}. 

Hierarchical time series forecasting is used because of the hierarchical structure formed by different generation technologies deployed in each region of the NEM. Both bottom-up and top-down procedures are used for forecasts. \cite{SW09} and \cite{HAA+11} presented the importance of forecast reconciliation in national account balancing and tourist demand, respectively. Due to the two-level hierarchy, we consider two forecast reconciliation methods, namely bottom-up and top-down, to reconcile the point forecast of electricity supplies by fuel mix, and improve forecast accuracy. The bottom-up approach entails generating forecasts for each disaggregated (or lower-level) series individually and then aggregating these forecasts to produce forecasts for the aggregated series \citep{Kahn98}. This method is particularly effective when the bottom-level series exhibit a high signal-to-noise ratio. The top-down method entails forecasting the completely aggregated series, and then disaggregating the forecasts based on historical proportions. \cite{GS90} introduced some possible ways of choosing these proportions. \cite{HAA+11} and \cite{PAG+21} show that top-down methods can introduce bias, even if the forecasts for the top level are unbiased.

In CoDa, we treat generation shares of each fuel type as compositional data, as the proportions of generation by fuel types are all non-negative and sum to $100\%$. Due to these constraints, we consider centered log-ratio and cumulative distribution function transformations to remove the constraints in Section~\ref{sec:4}.

Using our data set, the bottom-up forecast reconciliation provides the most accurate forecasts. The CoDa shares some similarity to the top-down forecast reconciliation method. Our results contribute to the literature by accurately predicting electricity supplies across various fuel mix types in five Australian states. The study will also benefit current participants, future investors and regulators in energy markets for trading, investment and system planning purposes. 

The remainder of the paper is presented as follows: In Section~\ref{sec:2}, we introduce the institutional background of the Australian NEM and the fuel mix data set. Sections~\ref{sec:3} and~\ref{sec:4} describe the two approaches, including the hierarchical time series forecasting and CoDa methods, and a comparison of the forecasting results. Section~\ref{sec:5} concludes, along with some ideas on how the methodology can be further extended.



\section{The fuel mix of the Australian electricity market}\label{sec:2}

\subsection{The Australian National Electricity Market}\label{sec:2.1}


The Australian NEM interconnects five regional market jurisdictions, including New South Wales (NSW), Queensland (QLD), South Australia (SA), Tasmania (TAS) and Victoria (VIC), and operates as a wholesale electricity market \citep{han2020}. The NEM is a spot market where demand and supply are balanced in real-time by the Australian Energy Market Operator (AEMO). It is also an `energy only' pool where all generated electricity has to be traded through this market as compared to many other power systems such as PJM and NYISO in the United States where a capacity market exists \citep{yan2020}. Trading in the NEM is through a competitive bidding and central dispatch process where power generators make offers to supply quantities of electricity at different prices for each 5-minute dispatch interval \citep{Han2025}. AEMO then determines the spot price for each state and the generators to be dispatched based on the market demand and a least-cost optimization, where generators with lower marginal costs are given priority. As generators in the same region are paid uniformly at the regional spot price for their supplied output, their trading strategies typically aim to maximize their supply in high-price periods.


Electricity demand exhibits a strong seasonal pattern driven by changes in underlying factors such as weather conditions and business or household activities \citep{Weron2006}. Specifically, demand is typically higher in summer and winter due to higher cooling or heating needs. Intra-weekly and intra-daily patterns can also be observed, with higher electricity usage on weekdays and during business operating hours.


The NEM leverages a diverse range of generation technologies for electricity supply, including fossil fuel generators, such as coal and gas plants, and renewable generators, such as wind and solar farms. According to the technical features of different generation technologies, their electricity supply fits for varied purposes and typically exhibits different patterns. In particular, coal-fired generators, due to their low operating cost and relatively low flexibility, tend to produce electricity continuously and serve as base-load generators. Gas-powered and hydroelectric generators are typically more flexible, with quicker ramp-up/down rates, making them suitable for peak load generation. They are used to supply both base-load and peak-load generation, depending on the local generation mix \citep{Nazifi2021}. For example, as the dominant generation resource in TAS, hydroelectric generators serve as base-load generators. In comparison, in NSW, QLD and VIC where coal is dominating, hydro plants are typically treated as peak-load generators. 

Gas and hydro generation, when providing peak load demand in the NEM, tends to be seasonal, peaking in summers and winters when electricity demand and prices are highest. Other peak-load generators in the NEM, such as battery storage and oil generators, are quickly responsive, but with limited capacity and expensive cost, they only operate during peak demand hours. Generations from VRE resources, including wind and solar, have low operating costs but a volatile generation profile due to their intermittent nature. Table~\ref{tab:1} lists local fuel types in each of the five states in the Australian NEM. Note that only grid-scale electricity supply is considered. Generation from consumer energy resources, such as rooftop solar generation, is considered a deduction from market demand.
\begin{table}[!htb]
\centering
\tabcolsep 0.2in
\caption{Electricity supply by various fuel types in each of the five states in Australia.}\label{tab:1}
\begin{tabular}{@{}lllll@{}}
\toprule
NSW & QLD & SA & TAS & VIC \\
\midrule
Battery Storage & Battery Storage & Battery Storage & Hydro & Battery Storage \\
Black Coal & Black Coal &  Diesel Oil & Natural Gas & Brown Coal \\
Diesel Oil & Methane & Natural Gas & Wind & Hydro \\
Hydro & Diesel Oil & Solar & &  Natural Gas \\
Natural Gas & Hydro & Wind & & Solar \\
Solar & Kerosene & & & Wind \\
Wind & Natural Gas & \\
	& Solar & \\
	& Wind & \\
\bottomrule
\end{tabular}
\end{table}

The Australian NEM is under a significant energy transition from traditional fossil fuels to renewable energy resources with the aim of reducing carbon emissions and achieving a net-zero emission economy by 2050 \citep{AEMO2024,Ming2024}. This transition has led to significant changes to the electricity supply and generation mix. Since 2013, over 6,000 MW of coal-fired generation assets have been removed from the market, and this gap has been filled by renewable generation mainly from wind and solar plants \citep{Regulator2023, Beggs2025}. The increasing share of VREs has introduced more volatility to market supply, making it more challenging for the NEM participants to optimize their trading and operating strategy, especially for flexible generators such as gas-powered generators and battery storage since they typically serve the residual demand after VREs. The transition has also resulted in additional challenges regarding the reliability of supply in the NEM \citep{fink17}, which is also evidenced by the June 2022 suspension of the wholesale electricity market \citep{simshauser2023,pourkhanali2024,Rangarajan2025}. Furthermore, for new investors entering the NEM, higher market volatility and increased difficulties with predicting short-term and long-term electricity demand, prices and fuel mix makes their investment return hard to estimate and leads to more uncertainties for decision-marking.

\subsection{Electricity supply data by fuel mix}

The data used in this study for the forecast of electricity generation fuel mix are daily electricity supply by different fuel technologies in the five regional electricity markets in the Australian NEM, namely, NSW, QLD, SA, TAS and VIC from January 1, 2019 to December 31, 2023. This data set is obtained from the database provided by AEMO \citep{AEMO_NEMWEB}.

\begin{figure}[!htb]
\begin{subfigure}[b]{0.5\textwidth}
\includegraphics[width=\textwidth, height=5.7cm]{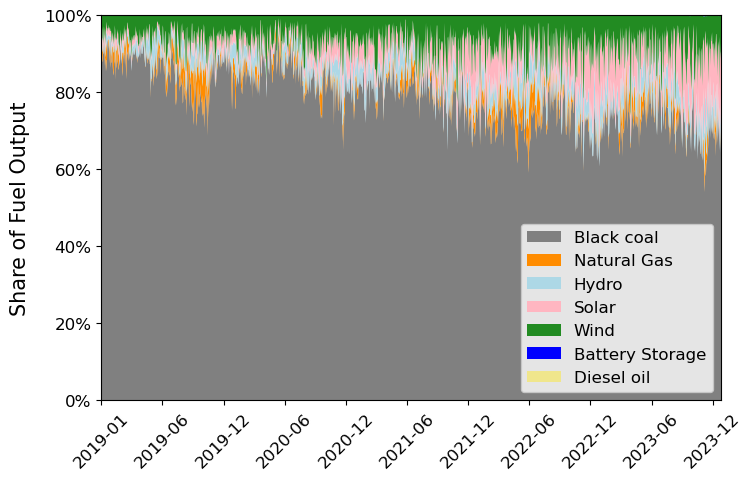}
\caption{NSW}\label{fig:fuel_ts_nsw}
\end{subfigure}
\begin{subfigure}[b]{0.5\textwidth}
\includegraphics[width=\textwidth, height=5.6cm]{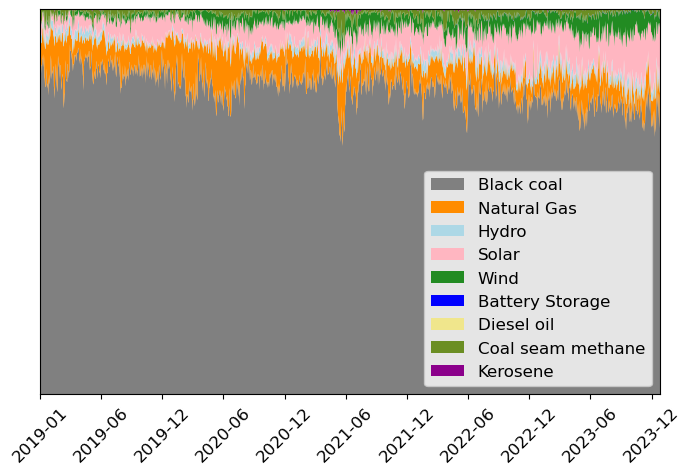}
\caption{QLD}\label{fig:fuel_ts_qld}
\end{subfigure}
\begin{subfigure}[b]{0.5\textwidth}
\includegraphics[width=\textwidth, height=5.7cm]{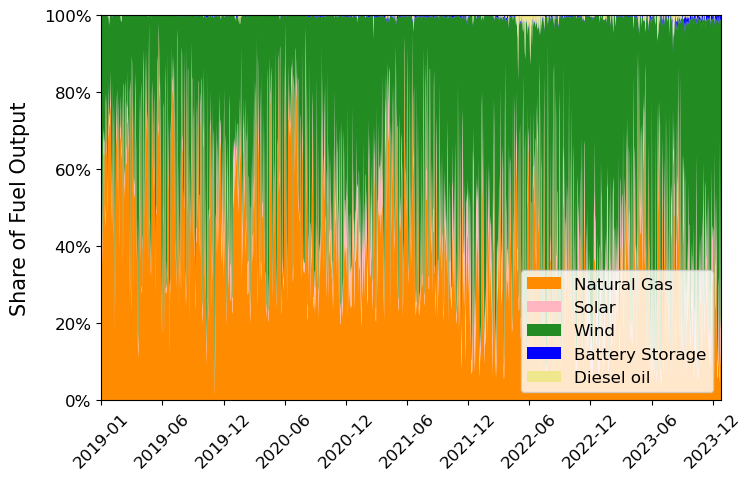}
\caption{SA}\label{fig:fuel_ts_sa}
\end{subfigure}
\begin{subfigure}[b]{0.5\textwidth}
\includegraphics[width=\textwidth, height=5.6cm]{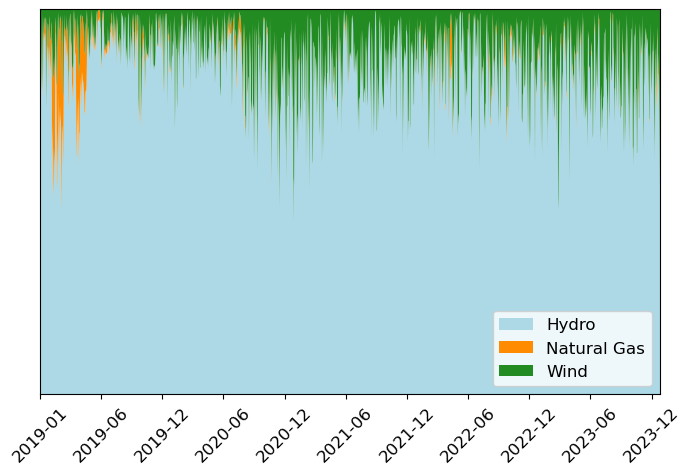}
\caption{TAS}\label{fig:fuel_ts_tas}
\end{subfigure}
\begin{subfigure}[b]{0.5\textwidth}
\includegraphics[width=\textwidth, height=5.7cm]{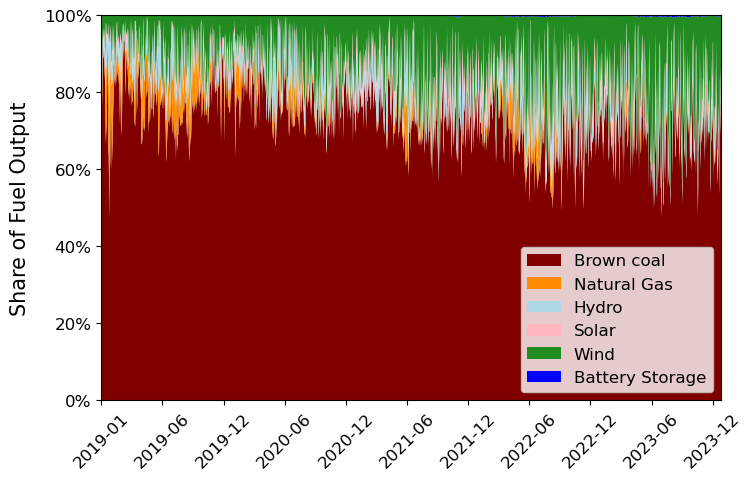}
\caption{VIC}\label{fig:fuel_ts_vic}
\end{subfigure}
\caption{\small{Generation mix over time for the NSW, QLD, SA, TAS and VIC states in Australia.}}\label{fig:fuel_ts}
\end{figure}

Figure~\ref{fig:fuel_ts} displays the daily electricity supply by different fuel types as proportions of the total generation in each region within the NEM. As depicted in the figure, the composition of the fuel mix exhibits fluctuations over time, with pronounced seasonality primarily driven by the seasonal availability of natural resources such as wind and solar energy. The figure also reveals a gradual trend indicative of the ongoing energy transition in recent years, characterized by an increasing share of renewable energy and a corresponding decline in fossil fuel usage. Moreover, there are notable regional variations in the generation fuel mix. Specifically, coal remains the predominant source of generation in NSW, QLD and VIC, while TAS relies heavily on hydropower, and SA is largely dependent on wind energy. Consequently, SA's electricity supply is marked by significant volatility due to the intermittent nature of wind power, whereas the generation mix in NSW and QLD is relatively more stable.

\section{Hierarchical time series forecasting}\label{sec:3}

\begin{figure}[!htbp]
\centering\begin{tikzpicture}
\tikzstyle{every node}=[minimum size = 18mm]
\tikzstyle[level distance=50cm] 
\tikzstyle[sibling distance=40cm]
\tikzstyle{level 3}=[sibling distance=46mm,font=\footnotesize]
\tikzstyle{level 2}=[sibling distance=42mm,font=\small]
\tikzstyle{level 1}=[level distance=30mm, sibling distance=26mm,font=\normalsize]
\node[circle,draw]{NSW}
   child {node[circle,draw] {Battery}
	  }
   child {node[circle,draw] {Coal}
		}
   child {node[circle,draw] {Oil}}
      child {node[circle,draw] {Hydro}}
         child {node[circle,draw] {Gas}}
            child {node[circle,draw] {Solar}}
   child {node[circle,draw] {Wind}
 };
\end{tikzpicture}
\caption{The hierarchy tree diagram, with seven fuel types of electricity consumptions in the state of NSW.}\label{fig:1}
\end{figure}
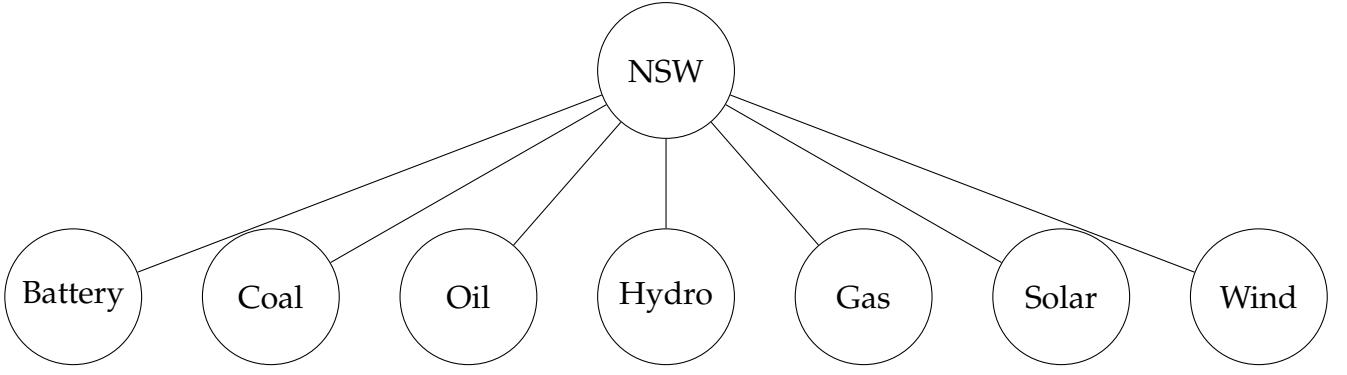

We consider the fuel mix type a disaggregation factor for modeling electricity demand in each state. For example, in NSW, the generation fuel mix contains seven fuel types. We denote all the base forecasts at day $t$ as $\bm{B}_t = (B_{1,t},\dots, B_{7,t})^{\top}$, where $^{\top}$ denotes matrix transpose, $t=1,\dots,n$ and $n$ denotes the total number of days. The hierarchical structure is shown in Figure~\ref{fig:1}.

We can express $\bm{T}_t=[T_t,\bm{B}_t]^{\top}$, a vector containing all series at both levels
\begin{small}
\begin{equation}
\underbrace{ \left[
\begin{array}{l}
T_{t} \\
B_{1,t}\\
B_{2,t} \\
B_{3,t} \\
B_{4,t} \\
B_{5,t} \\
B_{6,t} \\
B_{7,t} \\
\end{array}
\right]}_{\bm{T}_t} = \underbrace{\left[
\begin{array}{ccccccc}
1 & 1 & 1 & 1 & 1 & 1 & 1 \\
1 & 0 & 0 & 0 & 0 & 0 & 0  \\
0 & 1 & 0 & 0 & 0 & 0 & 0  \\
0 & 0 & 1 & 0 & 0 & 0 & 0  \\
0 & 0 & 0 & 1 & 0 & 0 & 0  \\
0 & 0 & 0 & 0 & 1 & 0 & 0  \\
0 & 0 & 0 & 0 & 0 & 1 & 0  \\
0 & 0 & 0 & 0 & 0 & 0 & 1  \\
\end{array}
\right]}_{\bm{\mathcal{S}}}
\underbrace{\left[
\begin{array}{l}
B_{1,t} \\
B_{2,t} \\
B_{3,t} \\
B_{4,t} \\
B_{5,t} \\
B_{6,t} \\
B_{7,t} \\
\end{array}
\right]}_{\bm{B}_t}, \label{eq:1}
\end{equation}
\end{small}
or $\bm{T}_t = \bm{\mathcal{S}}\bm{B}_t$, where $\bm{\mathcal{S}}$ is a summing matrix whose elements present an aggregation.

\subsection{Bottom-up procedure}

The bottom-up procedure has been widely used to reconcile forecasts in hierarchical structures \citep[see, e.g.,][]{DM92, ZT00}. The procedure aggregates the forecasts at the bottom level upwards towards the top level. Using the exponential smoothing of \cite{HKO+08}, these bottom-level forecasts are denoted by $\widehat{B}_{n+h|n} = [\widehat{B}_{1,n+h|n},\dots,\widehat{B}_{7,n+h|n}]^{\top}$, in the example of NSW. We then proceed to obtain reconciled forecasts $\widehat{\bm{T}}_{n+h|n}$ for all series as
\begin{equation*}
\widehat{\bm{T}}_{n+h|n} = \bm{\mathcal{S}}\widehat{\bm{B}}_{n+h|n},
\end{equation*}
where $\bm{\mathcal{S}}$ is the summing matrix in~\eqref{eq:1}.

\subsection{Top-down procedure}

The top-down procedure generates forecasts for the total series and then disaggregates these down the hierarchy. In the example of NSW, let $p_j$ for $j=1,2,\dots,7$ be a set of disaggregation proportions, where the forecast of the total series $T_{n+h}$ is disaggregated to the series at the bottom-level of the hierarchy. The disaggregation proportions can be computed from the historical averages \citep{GS90}, defined as
\begin{equation}
p_j^{\text{TDGSA}} = \frac{1}{n}\sum^{n}_{t=1}\frac{B_{j,t}}{T_t},\label{eq:2}
\end{equation}
where $j=1,\dots,7$. Each proportion $p_j$ represents the average of the ratios between the bottom-level series $B_{j,t}$ and the total series $T_t$. In contrast to~\eqref{eq:2}, one can compute the ratio of the averages between the bottom-level series and the total series
\begin{equation}
p_j^{\text{TDGSF}} = \frac{\sum^n_{t=1}B_{j,t}}{\sum^n_{t=1}T_t}. \label{eq:3}
\end{equation}
While~\eqref{eq:2} and~\eqref{eq:3} are deterministic, \cite{AAH09} introduced proportions based on forecasts rather than historical data, 
\begin{equation*}
p_j^{\text{TDFP}} = \frac{\widehat{B}_{j,n+h|n}}{\sum_{j=1}^{7}\widehat{B}_{j,n+h|n}},
\end{equation*}
where $\widehat{B}_{j,n+h|n}$ denotes the $h$-step-ahead forecasts for the bottom-level series.

\section{Compositional data analysis}\label{sec:4}

Generation by fuel mix relates strongly to the probability density function. For each day $t$, the generation proportions by fuel types are non-negative and sum to $100\%$. Because of the two constraints, the generation proportion can be viewed as compositional data. The sample space of compositional data is a simplex
\begin{equation*}
\mathbb{S}^{D-1} = \left\{(d_{t,1},d_{t,2},\dots,d_{t,D})^{\top}, \quad d_{t,x}\geq 0, \quad \sum^{D}_{x=1}d_{t,x} = 1\right\},
\end{equation*}
where $d_{t,x}$ denotes electricity generation for a given fuel mix $x$ in year $t$ in a state, $D$ denotes the total number of fuel mix types. Via a suitable transformation, the $D$-part compositional data are mapped from the simplex into a $(D-1)$-dimensional real space.

\subsection{Centered log-ratio (CLR) transformation}\label{sec:4.1}

In CoDa, \cite{AS80} and \cite{Aitchison82, Aitchison86} transform the compositional data from the simplex to Euclidean space using the CLR transformation:
\begin{equation}
\bm{s}_t = \{s_{t,x}\}_{x=1,\dots,D}=\ln\left\{\frac{d_{t,x}}{(\prod^{D}_{x=1}d_{t,x})^{1/D}}\right\}_{x=1,\dots,D}, \label{eq:1b}
\end{equation}
where the denominator of~\eqref{eq:1b} represents the geometric mean, playing the role of normalization. The unconstrained data matrix is denoted as $\bm{S} = (\bm{s}_1,\dots,\bm{s}_n)^{\top}$ with elements $s_{t,x}$ residing in real-valued space.

We apply principal component analysis to the data matrix $\bm{S}$, 
\begin{equation*}
s_{t,x} = \sum^{L}_{\ell=1}\widehat{\beta}_{t,\ell}\widehat{\phi}_{\ell,x}+\omega_{t,x},
\end{equation*}
where $\omega_{t,x}$ denotes model residual term for fuel type $x$ in day $t$, $\{\widehat{\phi}_{1,x},\dots,\widehat{\phi}_{L,x}\}$ and $\{\widehat{\beta}_{t,1},\dots,\widehat{\beta}_{t,L}\}$ denote the first~$L$ sets of estimated principal components and their scores for fuel type $x$ and day~$t$.

Using exponential smoothing, the one-step-ahead forecast of the $\ell$\textsuperscript{th} principal component score $\beta_{n+h,\ell}$ can be obtained. We use an automatic algorithm of \cite{HK08} to determine the optimal order of the exponential smoothing model based on the corrected Akaike information criterion. Conditioning on the estimated principal components and observed data, the forecast of $s_{n+h,x}$ can be obtained by
\begin{equation*}
\widehat{s}_{n+h|n,x} = \sum^L_{\ell=1}\widehat{\beta}_{n+h|n,\ell}\widehat{\phi}_{\ell,x}.
\end{equation*}

Via the inverse CLR transformation, the $h$-step-ahead forecast is given by
\begin{equation*}
\bm{\widehat{d}}_{n+h|n} = \Big[\frac{\exp^{\widehat{s}_{n+h|n,1}}}{\sum^D_{x=1}\exp^{\widehat{s}_{n+h|n,x}}},\dots,\frac{\exp^{\widehat{s}_{n+h|n,D}}}{\sum^D_{x=1}\exp^{\widehat{s}_{n+h|n,x}}}\Big].
\end{equation*}

\subsection{Cumulative distribution function (CDF) transformation}\label{sec:4.2}

Because of the logarithm operator, the CLR transformation can not handle zero values. As an alternative, \cite{SH25} propose a CDF transformation. It can be described below:
\begin{asparaenum}
\item[1)] Compute the empirical CDF via cumulative sum 
\[
S_{t,x} = \sum^{x}_{i=1}d_{t,i}, \quad x=1,\dots,D, \quad t=1,\dots,n, 
\]
where $S_{t,D}=1$.
\item[2)] We implement the logistic transformation,
\begin{equation*}
Z_{t,x} = \text{logit}(S_{t,x}) = \ln\left(\frac{S_{t,x}}{1-S_{t,x}}\right),
\end{equation*}
where $\ln(\cdot)$ denotes the natural logarithm.
\item[3)] We perform eigen-decomposition on the covariance of $Z_{t,x}$, 
\begin{equation*}
Z_{t,x} = \sum^{K}_{k=1}\widehat{\eta}_{t,k}\widehat{\psi}_{k,x}+\epsilon_{t,x},
\end{equation*}
where $\psi_{k,x}$ represents the $k$\textsuperscript{th} orthonormal principal component for a fuel mix $x$, $\eta_{t,k}$ represents the estimated principal component score at day $t$, $\epsilon_{t,x}$ represents residual function for a fuel mix $x$ in day~$t$, and $K$ denotes the number of retained principal components. We consider an eigenvalue ratio (EVR) criterion  of \cite{LRS20} to select the optimal value of $K$, and it can be expressed as
\begin{equation*}
K = \argmin_{1\leq k\leq n}\left\{\frac{\widehat{\lambda}_{k+1}}{\widehat{\lambda}_{k}}\times \mathds{1}\Big(\frac{\widehat{\lambda}_k}{\widehat{\lambda}_1}\geq \delta\Big) + \mathds{1}\Big(\frac{\widehat{\lambda}_k}{\widehat{\lambda}_1}<\delta\Big)\right\},
\end{equation*}
where $\widehat{\lambda}_k$ denotes the $k$\textsuperscript{th} estimated eigenvalue, tuning parameter $\delta$ is a pre-specified number, customarily $\delta = 0.001$, and $\mathds{1}(\cdot)$ is the binary indicator function.
\item[4)] By conditioning on the estimated functional principal components and observed data, the $h$-step-ahead forecast of $Z_{n+h,x}$ can be obtained
$\widehat{Z}_{n+h|n,x} = \sum^K_{k=1}\widehat{\eta}_{n+h|n,k}\widehat{\psi}_{k,x}$, where $\widehat{\eta}_{n+h|n,k}$ denotes the $h$-step-ahead univariate time-series forecasts of the $k$\textsuperscript{th} scores. 
\item[5)] By taking the inverse logit transformation, we obtain
\begin{equation*}
\widehat{S}_{n+h|n,x} =  \frac{\exp^{\widehat{Z}_{n+h|n,x}}}{1+\exp^{\widehat{Z}_{n+h|n,x}}}.
\end{equation*}
\item[6)] Via the first-order differencing $\Delta$, the $h$-step-ahead forecast is given by 
\begin{equation*}
\widehat{d}_{n+h|n,x}=\Delta_{i=1}^x\widehat{S}_{n+h|n, i}, 
\end{equation*}
where $\widehat{d}_{n+h|n,1} = \widehat{S}_{n+h|n,1}$. 
\end{asparaenum}

\subsection{Result}

We consider the problem of forecasting electricity fuel mix by different fuel types in all states of the Australian NEM. We divide the entire data sample from January 1, 2019 to December 31, 2023 into a training sample and a testing sample. While the training sample contains the initial 75\% of the total number of observations, the testing sample contains the remaining 25\%. Averaged over 457 days in the testing sample, the bottom-up procedure produces the smallest error. The error metric we considered is mean absolute scaled error (MASE), defined as
\begin{equation*}
\text{MASE}_{\gamma} = \frac{\frac{1}{D}\sum_{x=1}^{D}\left|d_{\gamma,x}-\widehat{d}_{\gamma|\gamma-1,x}\right|}{\frac{1}{D}\sum_{x=1}^{D}\left|d_{\gamma,x} - d_{\gamma-1,x}\right|}, \qquad
\overline{\text{MASE}} = \frac{1}{n_{\text{test}}}\sum^{n_{\text{test}}}_{\gamma=1}\text{MASE}_{\gamma}, 
\end{equation*}
where $d_{\gamma,x}$ denotes the $x$\textsuperscript{th} fuel mix type in the holdout data consisting of the $\gamma$\textsuperscript{th} day in the testing period and $\widehat{d}_{\gamma|\gamma-1,x}$ denotes the one-step-ahead forecast. The denominator is the mean absolute difference of the one-step-ahead na\"{i}ve forecasts, with $\bm{d}_{0}$ denoting the last year of the training data.
 
\begin{figure}[!htb]
\centering
\includegraphics[width=8.5cm]{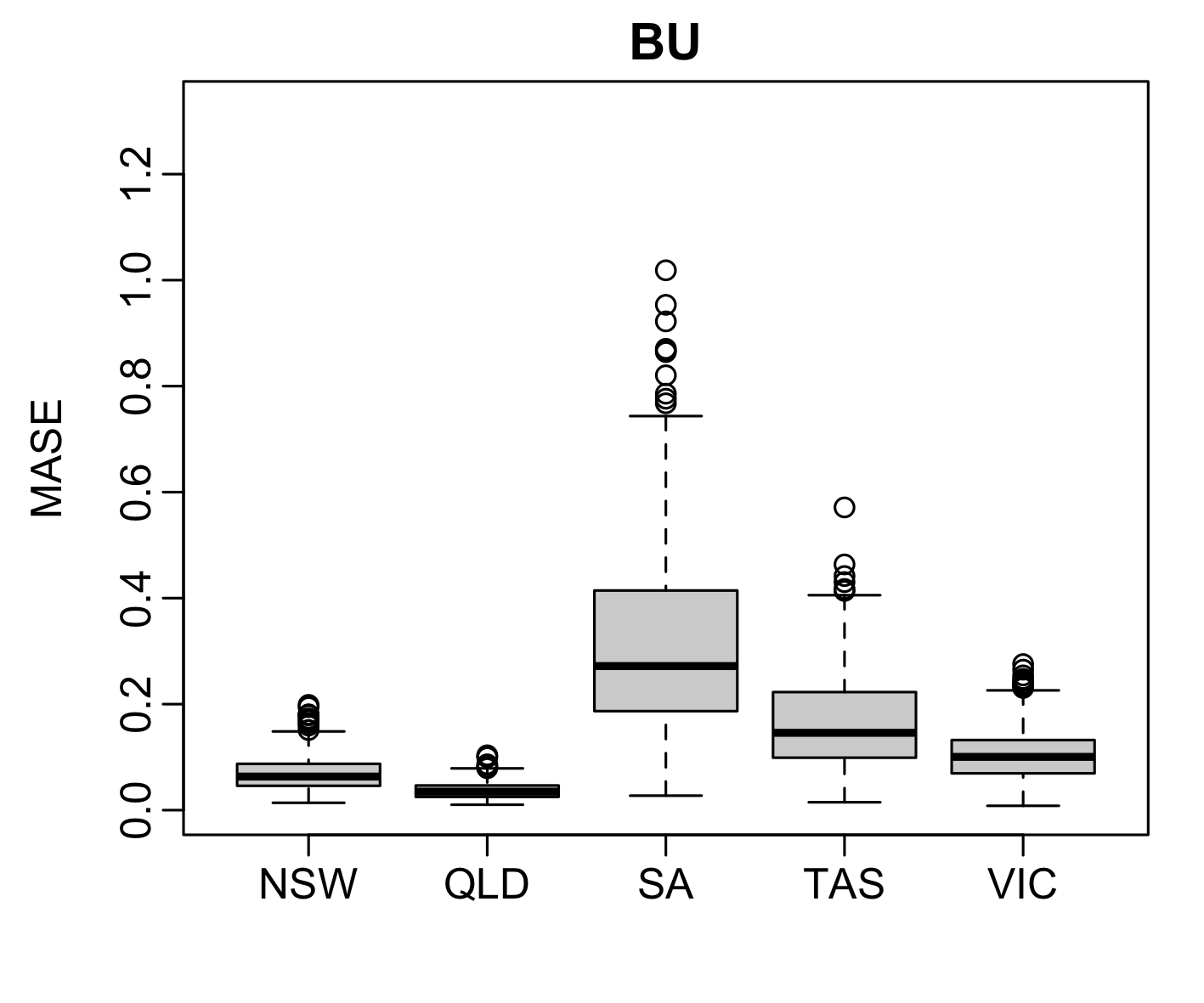}
\quad
\includegraphics[width=8.5cm]{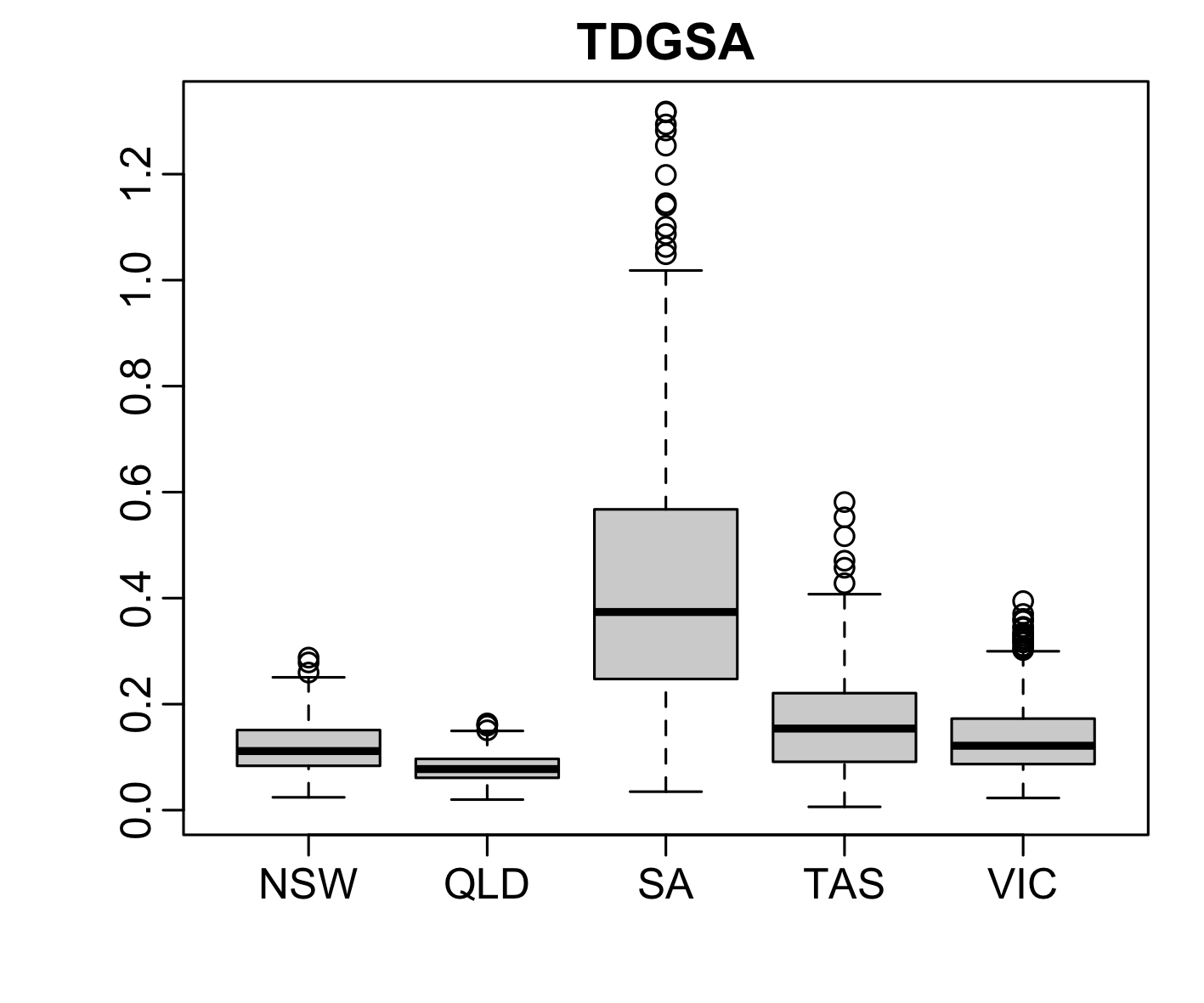}
\\
\includegraphics[width=8.5cm]{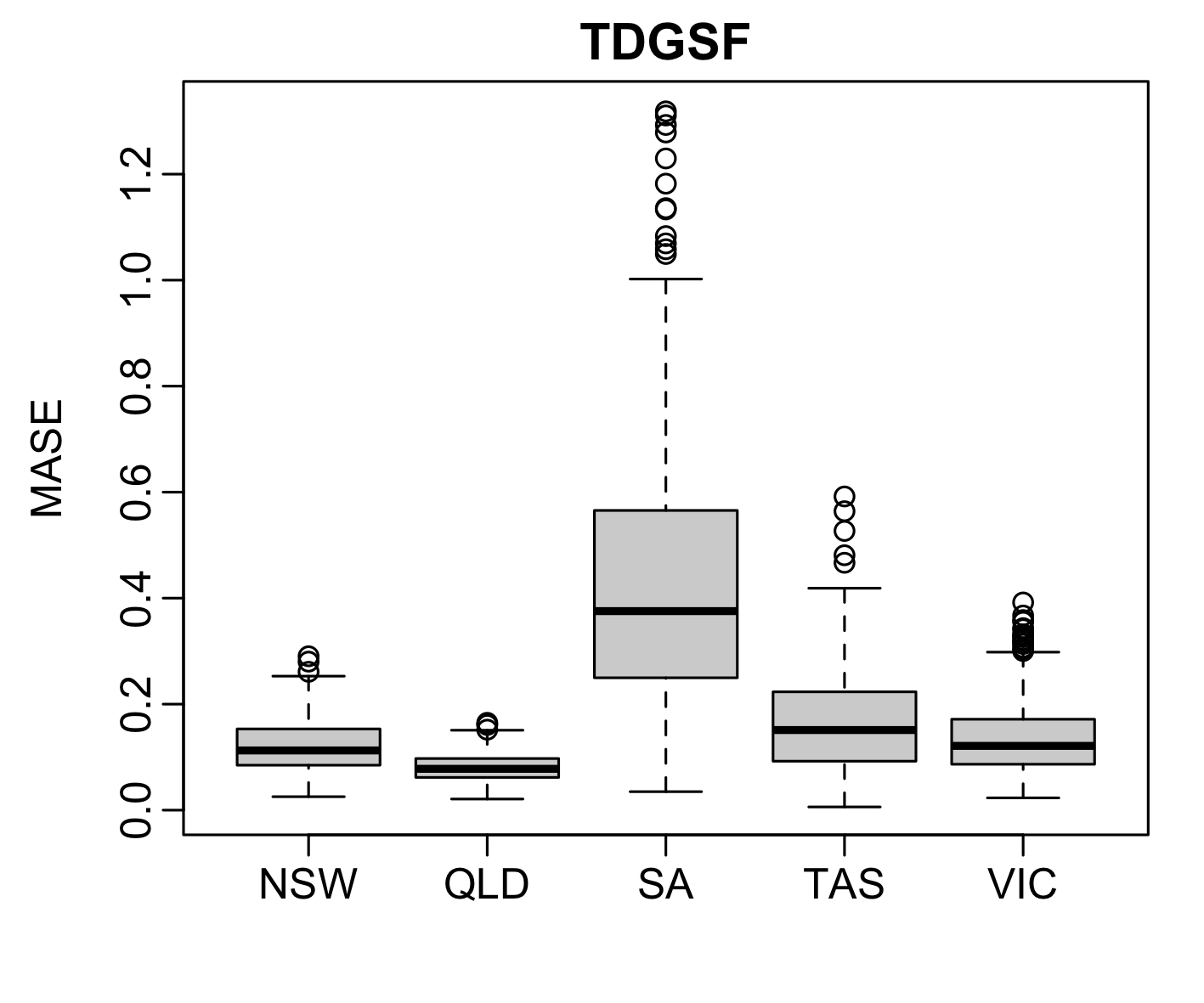}
\quad
\includegraphics[width=8.5cm]{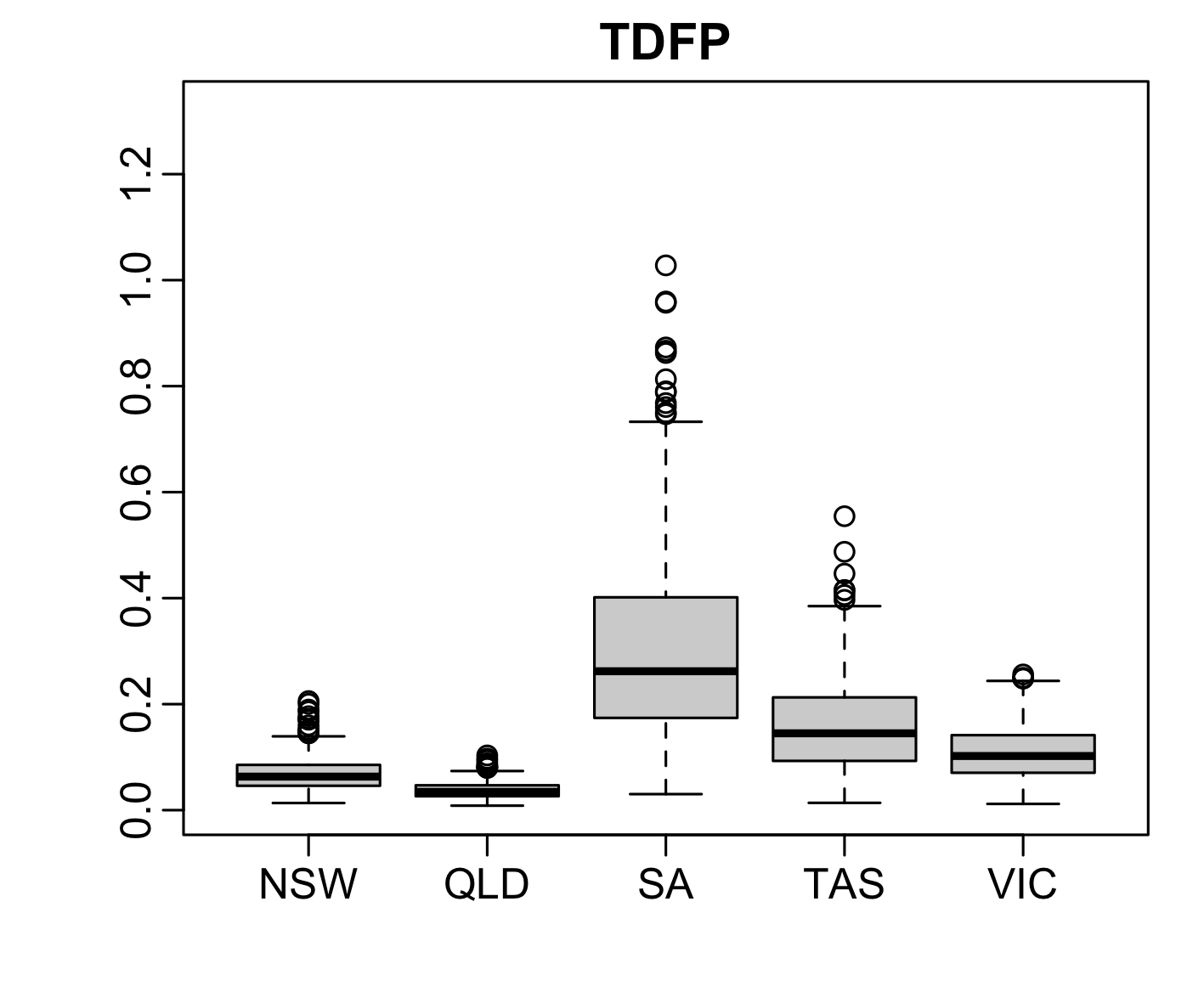}
\\
\includegraphics[width=8.5cm]{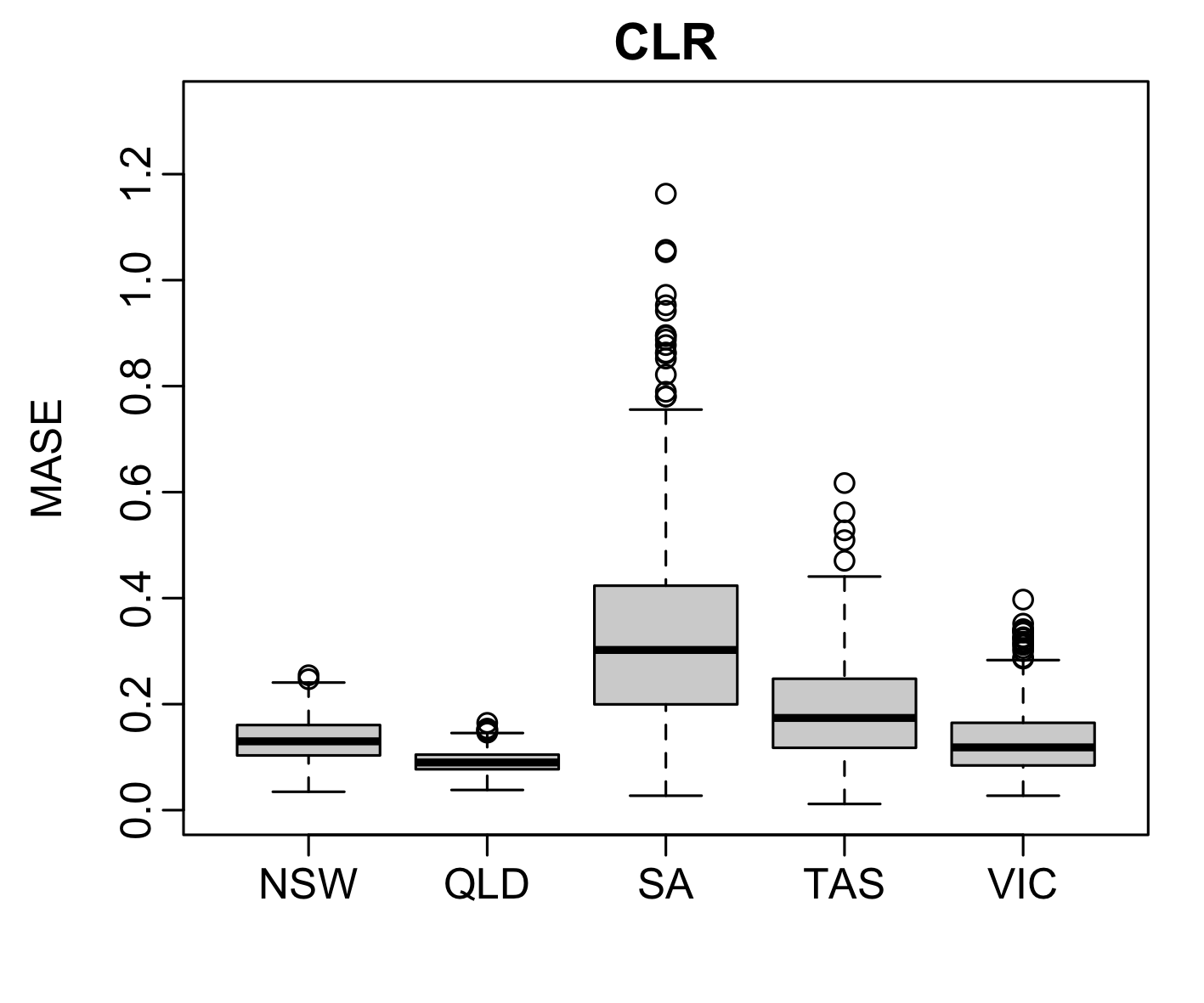}
\quad
\includegraphics[width=8.5cm]{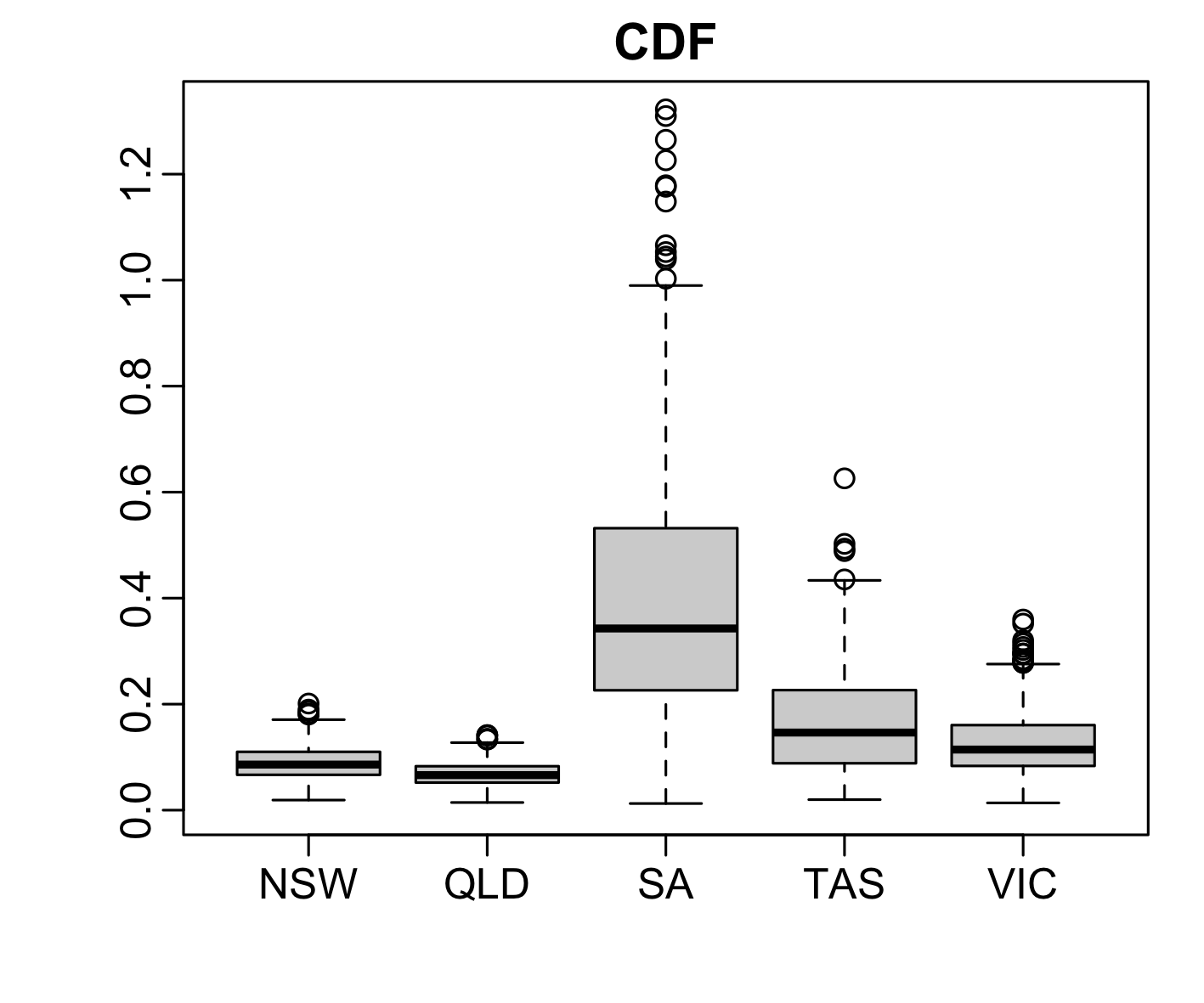}
\caption{Boxplots of MASE obtained from the four hierarchical time series methods and two CoDa transformations for forecasting one-day-ahead electricity supply in all five states of Australia by different fuel mix types.}\label{fig:2}
\end{figure}

In Figure~\ref{fig:2}, we display boxplots of MASE obtained from the four hierarchical time series methods and two CoDa transformations for modeling and forecasting one-day-ahead electricity supply by different fuel mix types. The bottom-up method is recommended across all five states. Across all methods, SA consistently exhibits the highest and most variable MASE, followed by TAS. Given the absence of coal generation in these markets, the high forecast inaccuracy is primarily attributable to the inherent variability of renewable energy sources, especially intermittent wind power, which dominates the energy mix in SA. Meanwhile, QLD, as the most coal-dependent state with the lowest share of renewable energy, achieves the highest forecast accuracy.

In Table~\ref{tab:2}, we present the averaged MASE over 457 days and the number of times in which one method outperforms the rest. For all five states, the hierarchical forecasting, particularly the bottom-up method, outperforms others with the smallest MASE, followed by the TDFP method. The bottom-up method generally produces the minimal MASE among the 457 days in the testing data set. Among the five states, SA presents the greatest challenge for accurate forecasting because of its high proportion of intermittent wind generation, which brings substantial volatility. In contrast, QLD demonstrates the highest forecast accuracy, as indicated by the lowest $\overline{\text{MASE}}$.
\begin{table}[!htb]
\centering
\tabcolsep 0.073in
\caption{\small{Among 457 days in the testing sample, we compute the averaged MASE among the four methods. In addition, we count the number of days where one method produces the smallest $\overline{\text{MASE}}$.}}\label{tab:2}
\begin{small}
\begin{tabular}{@{}lllllll|rrrrrr@{}}
\toprule
&\multicolumn{6}{c}{$\overline{\text{MASE}}$} & \multicolumn{6}{c}{Number of days} \\ \midrule
State & BU & TDGSA & TDGSF & TDFP & CLR & CDF & BU & TDGSA & TDGSF & TDFP & CLR & CDF \\
\midrule
NSW & \textBF{0.0689} & 0.1181 & 0.1197 & 0.0694 & 0.1327 & 0.0894 &   174 &    31 &    25 &   168 &     1 &    58 \\ 
 QLD & \textBF{0.0370} & 0.0797 & 0.0804 & 0.0373 & 0.0923 & 0.0684 &   204 &    13 &     7 &   203 &    0 & 30  \\ 
    SA & 0.3126 & 0.4209 & 0.4210 & \textBF{0.3049} & 0.3402 & 0.3975 &   127 &    30 &     4 &   126 &    55 &   115 \\ 
  TAS & 0.1657 & 0.1657 & 0.1659 & \textBF{0.1590} & 0.1920 & 0.1661 &   112 &    55 &    58 &   119 &    31 &    82 \\ 
  VIC & \textBF{0.1061} & 0.1378 & 0.1369 & 0.1090 & 0.1330 & 0.1281 &   157 &    41 &    33 &   134 &    54 &    38 \\ 
 \bottomrule
\end{tabular}
\end{small}
\end{table}

\section{Conclusion}\label{sec:5}

We forecast the generation fuel mix in the Australian NEM. Fuel mix forecasting plays a vital role in the efficient operation, reliability, and price discovery of wholesale electricity markets, as it informs dispatch decisions and reflects the varying short-run marginal costs of different generation technologies. As the share of intermittent renewables increases and dispatchable fossil-fuel generation declines, forecasting has become more complex, with greater uncertainty impacting system reliability, price expectations, and risk management strategies for both variable and flexible generators.

The NEM has a varying mix of resources for electricity generation in its different regional markets. This variability across regions leads to a hierarchical structure in the NEM's generation mix, where different regions contribute to the overall fuel mix. Additionally, the data on the generation fuel mix is compositional, with the proportions of generation by fuel type being non-negative and summing to one. To address these characteristics, we evaluate and compare two forecasting approaches: hierarchical time series forecasting and CoDa. For hierarchical time series forecasting, we assess both bottom-up and top-down procedures, exploring how each method captures the fuel mix's structure and dynamics across regions. 

Using the MASE as the evaluation metric, our analysis reveals that the bottom-up method of hierarchical forecasting outperforms all other methods tested, with the top-down method based on forecast proportions performing closely behind. Additionally, we find that fuel mix forecasting yields the most accurate results in markets where traditional fossil fuel power, particularly coal, dominates the generation portfolio. In contrast, in markets where renewable energy sources, especially intermittent VRE sources, are predominant, neither the hierarchical forecasting nor the compositional forecasting approach achieves optimal accuracy. This underscores the forecasting challenges posed by the growing integration of VRE resources. While beyond the scope of this study, future models could benefit from including specific meteorological variables, such as wind speed and solar radiation, to improve the accuracy of VRE generation forecasts.

The results provide valuable insights for market operators, investors, market participants, and regulators interested in electricity supply and demand forecasting. The renewable transition over the last two decades has increased the uncertainty and complexity of short-term operational planning, as reflected in recent reliability challenges observed in wholesale electricity markets \citep{Csereklyei2021, simshauser2023, Rangarajan2025}. Fuel mix forecasting is essential for the reliable and efficient operation of electricity markets, guiding real-time dispatch, system balancing, and reserve allocation decisions. 

Predictions on the generation mix enable grid operators to anticipate fluctuations and proactively schedule backup generation, demand response measures, or energy storage to maintain power system security and reliability. More accurate short-term fuel mix forecasts are also important for price discovery and risk management in the wholesale electricity market, allowing generators to better anticipate higher or lower renewable generation periods and adjust their strategies accordingly. Retailers and traders can leverage these forecasts to predict price movements, plan procurement, and develop hedging strategies that minimize costs and reduce exposure to market volatility. In the longer term, fuel mix forecasts provide investors with a clearer understanding of the operational efficiency and profitability of various generation technologies under different market conditions, guiding informed investment decisions. For regulators and policymakers, insights from fuel mix forecasting are beneficial in assessing market design, promoting renewable energy investments, and ensuring a reliable, low-emission energy portfolio. Overall, our results illustrate the benefits of forecast reconciliation techniques for the prediction of short-term electricity supply by different types of generation technologies and their use for system operators and market participants. For the ease of reproducibility, the computer \Rlogo \ code is readily available at \url{https://github.com/hanshang/Australian_electricity_supply_fuel_mix}.

The method used in this study is not limited to forecasting energy generation mix; it can also be extended to other time series applications involving the forecasting of multiple related time series organized in a hierarchical or grouped structure, such as forecasting sales in a supply chain or sectoral carbon emissions. There are several ways in which the methodology presented can be further extended in future work. For example, the current hierarchy structure is the simplest two-layer structure. As the structure becomes complex with more layers, it is possible to consider other forecast reconciliation methods, such as the middle-out and optimal combination methods in \cite{HAA+11}. Furthermore, the centered log-ratio transformation can not handle zero values. Another natural remedy is considering the $\alpha$-transformation of \cite{TPW16}, which resembles the Box-Cox transformation.

\newpage
\bibliographystyle{agsm}
\bibliography{fuel_mix.bib}

\end{document}